 \documentclass[aps,prl,preprint,superscriptaddress,showpacs,preprintnumbers,amsmath,amssymb]{revtex4}

\usepackage{graphicx}
\usepackage{dcolumn} 
\usepackage{bm}      
\usepackage{verbatim}%

\begin{document}


\title
{ Formation of an Icosahedral Structure during the Freezing of Gold
Nanoclusters: Surface-Induced Mechanism
}  



\author{ H.-S. Nam }
\email{nampo@plaza.snu.ac.kr}

\affiliation
{
School of Materials Science and Engineering, Seoul National University, Seoul 151-742, Korea
}

\author{ Nong M. Hwang }

\affiliation
{
Center for Microstructure Science of Materials, School of Materials Science and Engineering,
Seoul National University, Seoul 151-742, South Korea
}
\affiliation
{
Korea Research Institute of Standards and Science,
Taejon 305-600, South Korea
}

\author{ B.D. Yu }

\affiliation
{
Department of Physics, University of Seoul, Seoul 130-743, South Korea
}

\author{ J.-K. Yoon }

\affiliation
{
School of Materials Science and Engineering, Seoul National University, Seoul 151-742, Korea
}


\date{\today}

\begin{abstract}

The freezing behavior of gold nanoclusters was studied
by employing molecular dynamics simulations 
based on the semi-empirical embedded-atom method. 
Investigations of the gold nanoclusters
revealed that, just after freezing, ordered nano-surfaces with
a fivefold symmetry were formed with interior atoms remaining
in the disordered state. Further lowering of temperatures
induced nano-crystallization of the interior atoms that
proceeded from the surface towards the core region,
finally leading to an icosahedral structure. These dynamic
processes explain why the icosahedral cluster structure is
dominantly formed in spite of its energetic metastability.

\end{abstract}

%
%
%
%
%



\maketitle

%
%
Nano sized metal clusters containing tens to thousands of atoms
have attracted great attention due to their possible applications
as catalysts and surface
nanostructures \cite{cluster,HeerJensen,MoriartyBinns}.
In particular, understanding and predicting the structural properties
and formation of clusters produced from the liquid state or
gas phase are major concerns in investigations of
the controlled growth of low dimensional structures. Unlike bulk
materials, small metal clusters exhibit various structural
modifications, for example, for fcc metals, cuboctahedra (CO) with
a face-centered-cubic (fcc) structure, twinned fcc (containing one
or several parallel twin planes) \cite{Pinto,Buffat}, twinned
hexagonal close packed (hcp) \cite{Yacaman}, icosahedral and
truncated icosahedral (Ih)
\cite{Buffat,MarksMartin,Ino,Komoda,Ascencio}, truncated
decahedral (Dh) \cite{MarksMartin,Ino,Komoda}, and amorphous
\cite{Garzon}. Notably, high-resolution electron microscopy
(HREM), under typical cluster-growth conditions, routinely
detected metal clusters of Ih or Dh morphology with a fivefold
symmetry of noncrystallographic atomic arrangements
\cite{MarksMartin}.

The thermodynamical-equilibrium forms and structures of metal
clusters have been exhaustively searched and discussed on the basis
of theoretical calculations. In contrast to experimental
observations, theoretical calculations suggest that for Au,
the Ih structure is energetically  metastable even for small
clusters (less than $100$ atoms) with a large surface-to-volume
ratio \cite{Cleveland,BalettoJCP}. With increasing cluster
size, the stability of the Ih structure should decrease markedly
due to accumulated strain energy \cite{Cleveland}. HREM studies
\cite{Buffat,MarksMartin,Ino,Komoda,Ascencio}, however, have revealed
that even larger clusters (up to a few thousand atoms) still
have Ih or Dh morphology. Real-time microscopic studies by
Iijima and Ichihashi \cite{Iijima} demonstrated structural changes in gold clusters
from a single crystalline form to a twinned crystalline form of the
Ih or Dh structures, and vice versa, originating from 
electron beam irradiation (charging effect). Such experimental
observations emphasize that the formation of Ih clusters is
governed by kinetic rather than thermodynamic factors.

In order to understand the kinetics, we investigated 
the structural changes of gold nanoclusters during cooling from a molten state 
by employing molecular dynamics (MD) simulations. 
Interestingly, we found that the Ih structure originates from surface ordering
not from conventional core nucleation. At the initial stage of the freezing, 
close-packed (111)-type planes with a fivefold Ih symmetry were formed at surfaces 
before the interior of the cluster crystallized. 
Subsequently, nano-crystallization at lower temperatures propagated inward 
from these (111)-type surface segments, leading to the Ih cluster.

All MD simulations of gold nanoclusters were performed 
using the semi-empirical embedded-atom method \cite{EAM}. 
We used the XMD code described in Ref.~\onlinecite{XMD}. 
In the simulations we used an Ih cluster of 561 atoms as an initial structure. 
The cluster was then heated to 1500 K, well above its melting temperature
to ensure complete melting, and equilibrated for a long period of more than 250 ps 
($\sim 1.3 \times 10^5 \Delta t$, where $\Delta t=$2.0 fs represents the time step 
for the integration of the equation of motion). 
The cluster was cooled to 300 K at $10^{11}$ K/s ($2 \times 10^{-4}$ K/$\Delta t$); 
the temperature was set through the mean kinetic energy of the atoms. 

\begin{figure}
\includegraphics[width=0.42\textwidth]{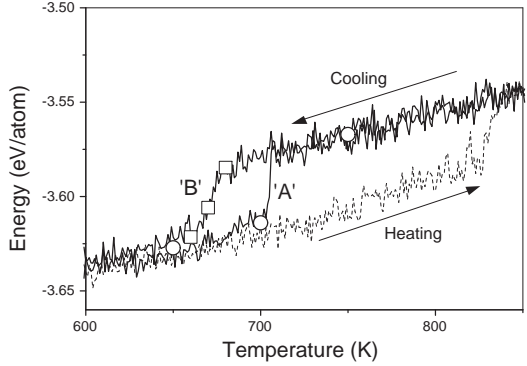} 
\caption
{ \label {fig:Epotential}
Variation of potential energy with temperature for gold clusters of 561 atoms 
with heating and cooling rates of $10^{11}$ K/s.
The dashed line denotes the melting curve during heating and the solid line
the freezing curve.
The different freezing curves `A' and `B' were obtained from different
initial configurations of melted clusters.
}
\end{figure}

Our simulations at different initial configurations showed that
most of the $561$ atom clusters were frozen to an Ih structure 
at a cooling rate of $10^{11}$ K/s. 
Typical melting and freezing behavior of the clusters is displayed 
by the potential energy vs. temperature curve in Fig.~\ref{fig:Epotential}. 
The melting and freezing phase transition can be identified 
by an abrupt change in potential energy. 
Interestingly, we here see two different types of freezing behavior 
under identical cooling condition.
In the first case shown by the freezing curve `A' in Fig.~\ref{fig:Epotential}, 
the cluster underwent a sharp liquid-solid transition and the freezing point was estimated as $706$ K.
In the second case shown by the freezing curve `B' in Fig.~\ref{fig:Epotential}, 
the liquid-solid transition took place over tens of degrees of temperature.

\begin{figure*}
\includegraphics[width=0.60\textwidth]{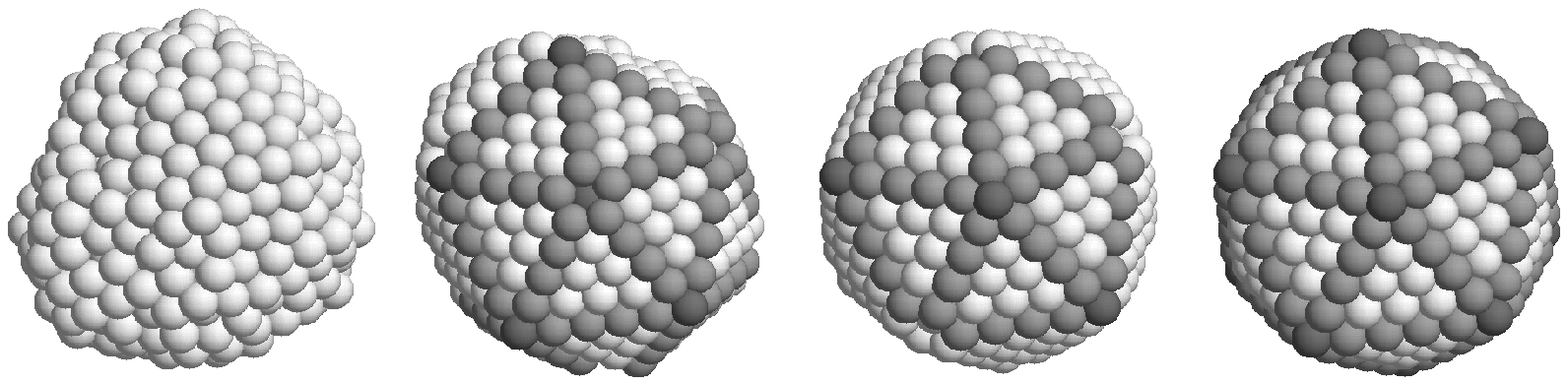} 
\includegraphics[width=0.60\textwidth]{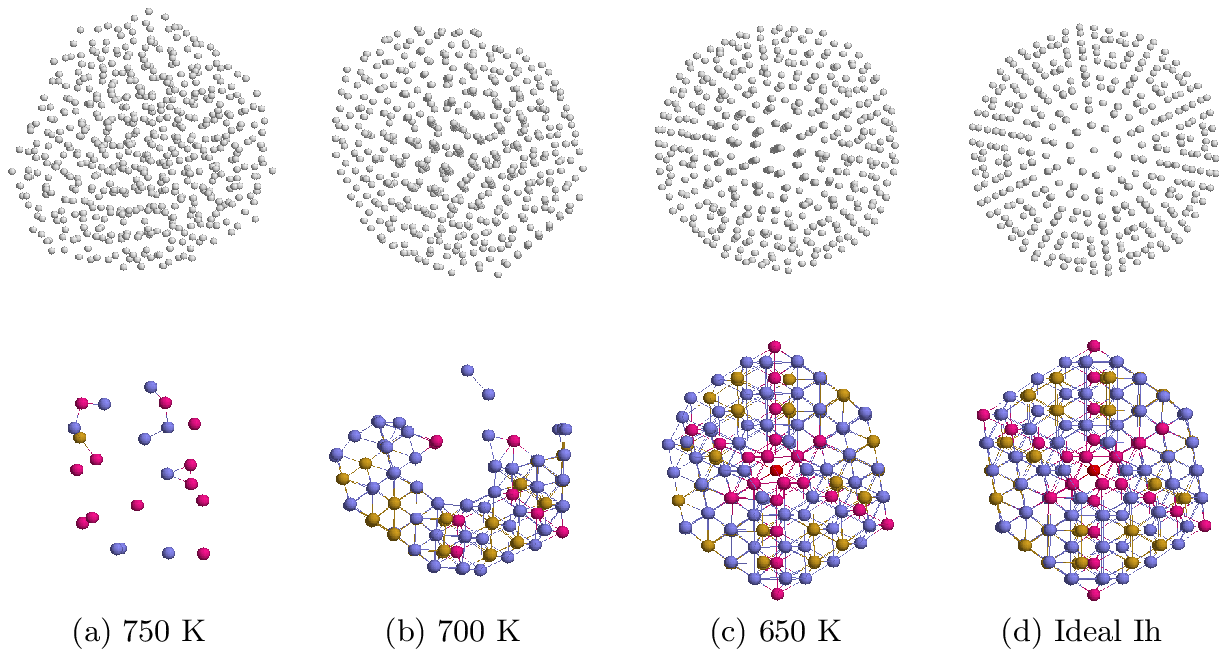} 
\caption
{ \label {fig:FirstCase}
(color) Cluster configurations at different stages
corresponding to the open circles in the freezing curve `A' of
Fig.~\ref{fig:Epotential}:
(a) in a liquid state (at $750$ K),  (b) just after freezing (at $700$  K),
(c) after complete rearrangement (at $650$  K) and (d) an ideal icosahedron for comparison.
In the upper row, only surface atoms are shown
while in the middle row, all the atoms are shown at a smaller size to display their inner arrangement.
In the lower row, solid-like atoms with a well-defined local symmetry are shown 
in two-cross-sectioned views by a ball-and-stick model.
Blue, gold, and red balls represent atoms with hcp, fcc, and fivefold local symmetries, respectively.
}
\end{figure*}

Similar MD simulation works of the melting and freezing 
were reported by Chushak et al. \cite{Chushak}  
and they also found the preferential formation of an Ih structure 
for various cluster sizes. In spite of the MD simulations, an understanding as to 
why the formation of an Ih structure occurs so frequently is still lacking. 
For detailed analysis of the freezing behavior 
we looked at cluster configurations corresponding to the three open circles 
(at $750$, $700$, and $650$ K, respectively) in freezing curve `A' of Fig.~\ref{fig:Epotential} 
[see Figs.~\ref{fig:FirstCase}(a), \ref{fig:FirstCase}(b) and \ref{fig:FirstCase}(c)]. 
Only for the purpose of avoiding vibrational noise effect 
in the analysis of the cluster structure, 
we relaxed the cluster-atomic configurations to the local energy-minimum structures 
by using the conjugate gradient minimization technique. 
We also showed an ideal Ih structure of a $561$-atom cluster 
for comparison [see Fig.~\ref{fig:FirstCase}(d)]. 
At $750$ K, which is well above the freezing temperature, the cluster was in a liquid state. 
Forming and dissolving of very small embryos occurred within the disordered state
and the cluster shape was highly fluctuating and approximately spherical. 
Notably, as the freezing temperature was approached, 
flat surface segments like solid facets started to appear temporarily. 
The abrupt decrease of potential energy shown in Fig.~\ref{fig:Epotential} 
indicates that the cluster at $700$ K had just frozen.
At 700 K, the cluster showed 
ordered facets with a fivefold symmetry, apparently with an Ih shape
[see Fig.~\ref{fig:FirstCase}(b)]. As the cluster was
cooled further to $650$ K, the interior atoms rearranged from these
surface facets to form an Ih crystalline structure [see
Fig.~\ref{fig:FirstCase}(c)], which was comparable to an ideal
icosahedron [see Fig.~\ref{fig:FirstCase}(d)].

The cluster configurations have been also investigated 
by using the local bond order parameter $q_l(i)$ 
characterizing the local atomic arrangement around atom $i$ \cite{Fuente}. 
The method is essentially the same in formalism as that of Ref.~\onlinecite{Chushak}, 
but local structure criteria are different. 
Thereby, we are successfully able to discriminate atoms with an hcp local symmetry for an Ih structure. 
Solid-like atoms with a well-defined local symmetry 
are shown in two-cross-sectioned views in the bottom row of Fig.~\ref{fig:FirstCase}. 
At high temperatures we observed typical forming and dissolving of very small embryos 
as expected [see Fig.~\ref{fig:FirstCase}(a)]. 
At 700 K, just after freezing, we clearly observed a surface-crystallized structure, 
representing that {\em the nanocrystallization for an Ih structure proceeds 
from surfaces segments towards core regions.} 

\begin{figure}
\includegraphics[width=0.42\textwidth]{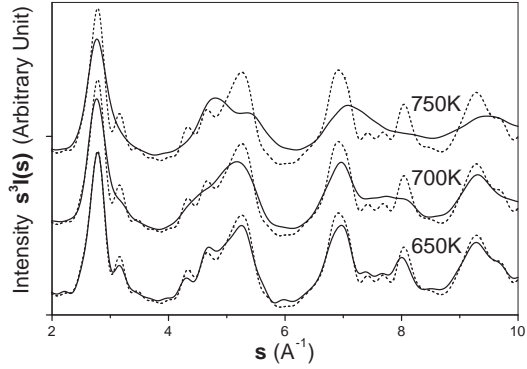} 
\caption
{\label{fig:diffraction}
Calculated diffraction patterns for the structures of
Figs.~\ref{fig:FirstCase}(a), \ref{fig:FirstCase}(b) and
\ref{fig:FirstCase}(c).
The dashed lines represent the reference diffraction pattern of
an ideal icosahedron of Fig.~\ref{fig:FirstCase}(d).
}
\end{figure}

Structure evolution can also be investigated using diffraction patterns. 
Figure~\ref{fig:diffraction} shows the simulated diffraction intensity $s^{3}I(s)$ 
versus the diffraction parameter $s$ for the clusters at $750$ , $700$,  and $650$  K
as compared to an ideal Ih cluster, denoted by a dashed line. 
Note that $s^{3}I(s)$ rather than $I(s)$ was used 
for detailed comparisons with the ideal Ih structure \cite{Waal}. 
At $750$ K, the diffraction pattern was somewhat broadened, 
due to the liquid state of the cluster [see Fig.~\ref{fig:FirstCase}(a)]. 
At $700$ K, some peaks started to appear, which represented a disordered structure 
with a short-range order [see Fig.~\ref{fig:FirstCase}(b)]. 
At $650$ K, fine structures of peaks were found and the overall pattern was similar
to that of the ideal Ih structure.

\begin{figure}
\includegraphics[width=0.35\textwidth]{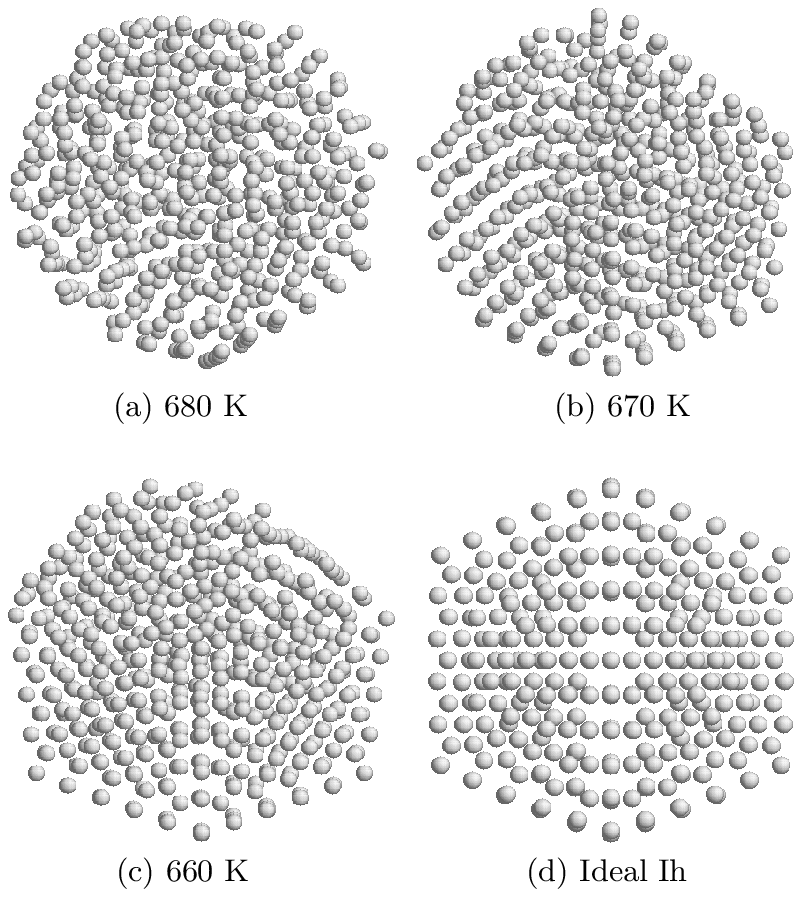}
\caption
{\label{fig:SecondCase}
Cluster configurations at different stages corresponding
to open squares in the freezing curve `B' of
Fig.~\ref{fig:Epotential}:
(a) of the initial stage at $680$ K, (b) of the middle stage at $670$ K,
(c) of the final stage at $660$  K and
(d) of an ideal icosahedron for comparison.
}
\end{figure}

Next we consider the second case of freezing curve `B' that
appears to be different from freezing curve `A' [see
Fig.~\ref{fig:Epotential}]. In this case, freezing took place
by the advancement of the solid/liquid interface relatively
gradually (over about 200 ps). This is in contrast to the freezing situation
described by curve `A' where freezing took place from a liquid
state to a surface-ordered structure abruptly (over about $10$ ps).
Figures ~\ref{fig:SecondCase}(a), ~\ref{fig:SecondCase}(b), and
~\ref{fig:SecondCase}(c) show cluster configurations corresponding
to the three open square symbols (at 680, 670, and 660 K,
respectively) in freezing curve `B' of
Fig.~\ref{fig:Epotential}. During the initial stage of freezing [see
Fig.~\ref{fig:SecondCase}(a)], facets
started to form on the local surface. In the middle stage [see
Fig.~\ref{fig:SecondCase}(b)], more surface facets were formed
giving the surface a fivefold symmetry, and crystallization
proceeded from one part of surface to the others. In the final
stage [see Fig.~\ref{fig:SecondCase}(c)], the entire cluster was
crystallized, although the final cluster was not a single crystal
but a multiply twined crystal.

According to classical nucleation theory 
based on the macroscopic concept of the interface \cite{Handbook,RMW}, 
nucleation should take place in the interior of the liquid cluster \cite{Bartell}.
In this case, the stable fcc phase is expected to be formed locally inside the cluster 
and to finally grow into a single fcc crystalline phase. 
Then, the final fcc structure would be energetically more stable than the Ih structure. 
However, our MD simulations revealed that the Ih structure is a prevailing cluster
structure, which is in agreement with experimental observations.
{\em Inspections of the freezing cluster configurations showed
that Ih cluster formation is initiated from surface
ordering not from conventional core nucleation.} 
In the freezing processes of the melted cluster shown in curve `A' of Fig.~\ref{fig:Epotential}, 
surface ordering took place over the entire surface. 
Surface ordering induces an abrupt transition from the liquid to an amorphous state. 
Actually, the potential energy of an amorphous cluster with an ordered surface is 
only slightly higher than that of the Ih cluster. 
In the case of curve 'B' in Fig.~\ref{fig:Epotential}, 
surface ordering took place only on parts of the local surface 
and propagated. 
Our results suggest that low kinetic barriers at surfaces make cluster formation 
of the meta-stable Ih skin structure kinetically favorable 
(although the energy barriers are not accessible under the present MD simulations). 
Once the ordered surface facets were formed, they would work as a crystallization seed, 
and thereby nano-crystallization would proceed inward, finally forming the Ih structure.

In our further simulations of different cooling conditions 
and cluster sizes up to $1000$ atoms, 
including cluster sizes of 581 atoms and of 545 atoms 
corresponding to the magic number sizes of the fcc octahedron 
and of the truncated decahedron, respectively, 
other structures such as the truncated decahedral and fcc structures were also formed, 
but the Ih structure was always dominant. 
Kinetic competition between these structures
seems to be correlated with the surface ordering of the cluster during freezing, 
which means that the reduction of surface energy more efficiently contributes 
to the determination of the cluster structure than the internal energy.
In many simulation runs, we found that when surface ordering was dominant,
the cluster became Ih. Otherwise, the cluster became t-Dh or fcc.

In our simulations, we investigated a structural change of clusters
during cooling from the liquid state. Typically, clusters are
generated from the gas phase through inert gas condensation using
adiabatic expansion. Even in this case, if clusters grow as a
liquid droplet and then solidify \cite{BalettoCPL}, or if
disordered or amorphous clusters are first formed and then these
undergo solid-solid transition to nanocrystalline structures,
surface ordering of the clusters might occur and have the same effect
as freezing.

In summary, by performing MD simulations on gold clusters,
we examined the formation mechanism of an Ih cluster experimentally observed.
When $561$-atom gold clusters were cooled from a liquid state,
the Ih structure was obtained repeatedly by two different processes.
In the first case, surface ordering with \{$111$\}-type facets with
a five-fold symmetry took place
while interior atoms were frozen into a disordered structure abruptly,
over about $10$ ps.
The rearrangement of interior atoms then proceeded inward from the Ih-like surface, 
leading to an Ih cluster.
In the second case, crystallization took place at a surface region
and then propagated to the whole over about $200$ ps.
In both cases, the formation of the Ih structure
originated from ordered surface facets with a fivefold symmetry.
This surface-induced mechanism explains why clusters tended to adopt an Ih structure,
although it is not energetically the most stable phase.


\begin{acknowledgments}
We gratefully acknowledge support from the Korea Ministry of Education 
through the Brain Korea $21$ Program (H.-S.N. and J.-K.Y.), 
the Korea Ministry of Science and Technology 
through the Creative Research Initiative Program (N.M.H.), 
and the Korea Science and Engineering Foundation 
through the ASSRC at Yonsei University (B.D.Y.).
Fruitful discussions with S. C. Lee at Korea Institute of Science and Technology 
and P. R. Cha at Seoul National University are appreciated.
\end{acknowledgments}




\end{document}